\documentstyle[psfig,aasms4,flushrt]{article}
\newcommand {\BS}{{{\it Beppo}SAX}\ \ignorespaces}
\def\gta{ \lower .75ex \hbox{$\sim$} \llap{\raise .27ex \hbox{$>$}} }
\def\lta{ \lower .75ex\hbox{$\sim$} \llap{\raise .27ex \hbox{$<$}} }
 
\begin{document}

\title{Testing  Comptonizing coronae on a long \BS
observation of the Seyfert 1 galaxy NGC 5548}

\author{P.O. Petrucci\altaffilmark{1,2}, F. Haardt\altaffilmark{2},
 L. Maraschi\altaffilmark{1}, P. Grandi\altaffilmark{3},
 G. Matt\altaffilmark{6}, F. Nicastro\altaffilmark{3,4,5},
 L. Piro\altaffilmark{3}, G.C. Perola\altaffilmark{6}, A. De
 Rosa\altaffilmark{3}.}

\altaffiltext{1}{Osservatorio Astronomico di Brera, Milano, Italy} 
\altaffiltext{2}{Universit\'a dell'Insubria, Como, Italy}
\altaffiltext{3}{IAS/CNR, Roma, Italy} 
\altaffiltext{4}{CfA, Cambridge Ma., USA}
\altaffiltext{5}{Osservatorio Astronomico di Roma, Roma, Italy}
\altaffiltext{6}{Universit\'a degli Studi ``Roma 3'', Roma, Italy}

\begin{abstract}
We test accurate models of Comptonization spectra over the high quality
data of the \BS long look at NGC 5548, allowing for different geometries
of the scattering region, different temperatures of the input soft photon
field and different viewing angles.  We find that the \BS data are well
represented by a plane parallel or hemispherical corona viewed at an
inclination angle of 30$^{\circ}$. For both geometries the best fit
temperature of the soft photons is close to 15$^{+3}_{-9}$ eV. The
corresponding best fit values of the hot plasma temperature and optical
depth are $kT_{\rm e}\simeq$ 250--260 keV and $\tau\simeq$ 0.16--0.37 for
the slab and hemisphere respectively.  These values are substantially
different from those derived fitting the data with a power-law + cut off
approximation to the Comptonization component ($kT_{\rm e}\lta$ 60 keV,
$\tau\simeq$ 2.4).  In particular the temperature of the hot electrons
estimated from Comptonization models is much larger.  This is due to the
fact that accurate Comptonization spectra in anisotropic geometries show
"intrinsic" curvature which reduces the necessity of a high energy
cut-off.  The Comptonization parameter derived for the slab model { is}
larger than predicted for a two phase plane parallel corona in energy
balance, suggesting that a more ``photon-starved'' geometry is
necessary. The case of a hemispheric corona is consistent with energy
balance but requires a large reflection component.  The spectral
softening detected during a flare which occurred in the central part of
the observation corresponds to a decrease of the Comptonization
parameter, probably associated with an increase of the soft photon
luminosity, the { hard} photon luminosity remaining constant. The
increased cooling fits in naturally with the derived decrease of the
coronal temperature $kT_{\rm e}$ in the high state.
\end{abstract}


\section{Introduction}
The X-ray emission of Seyfert galaxies is commonly believed to be
produced by Compton scattering of soft photons on a population of hot
electrons { (Pozdniakov et al., \cite{poz76}; Shapiro et al.,
\cite{sha76}; Liang \& Price, \cite{lia77}; Sunyaev \& Titarchuk,
\cite{sun80})}. Although high energy observations of Seyfert galaxies may
be consistent with both thermal and non--thermal models (Zdziarski et al.
\cite{zdz94}; Malzac et al. \cite{mal98}) the non--detection of Seyferts
by Comptel and the high energy cut-offs indicated by OSSE (Jourdain et
al. \cite{jou92}; Maisack et al., \cite{mai93}) and {\it Beppo}SAX (Matt
\cite{mat99}) have focused attention on thermal models. In fact it was
shown early on that assuming thermal equilibrium and a plane parallel
geometry for the Comptonizing plasma above an accretion disk, and
neglecting direct heating of the disk, the average properties of the
X-ray emission of Seyfert galaxies could be naturally accounted for
(Haardt \& Maraschi \cite{haa91}).  The spectral slope is mainly
determined by two parameters, the temperature $kT_{\rm e}$ and optical
depth $\tau$ of the scattering electrons, while the cut--off energy is
related essentially to $kT_e$. Thus, simultaneous measurements of the
slope of the X-ray continuum {\it and} of the cut--off energy are
necessary to determine the physical parameters of the Comptonizing
region.

Moreover, in a disk plus corona system, the Comptonizing region and the
source of soft photons are {\it coupled}, as the optically thick disk
necessarily reprocesses and reemits part of the Comptonized flux as soft
photons which are the seeds for Comptonization.  The system must then
satisfy equilibrium energy balance equations, which depend on {\it
geometry} and on the ratio of direct heating of the disk to that of the
corona. In the limiting case of a "passive" disk, the amplification of
the Comptonization process, determined by the Compton parameter
$\displaystyle y\simeq 4\left(\frac{kT_e}{m_ec^2}\right)\,
\left[1+4\left(\frac{kT_e}{m_ec^2}\right)\right] \tau (1+\tau)$, is fixed
by geometry only. Therefore, if the corona is in energy balance, the
temperature and optical depth must satisfy a relation which can be
computed for different geometries of the disk+corona configuration (e.g.,
the review of Svensson \cite{sve96}). It is then theoretically possible
to constrain the geometry of the system and verify the selfconsistency of
the model, provided that $kT_e$ and $\tau$ are known with sufficient
precision.

The extent of spectral variability during luminosity variations is an
additional, in principle powerful, diagnostic tool that can be used to
test existing models, as it provides direct insight into the way the
emitting particles are heated and cooled. For example, if the plasma is
pair dominated, the value of $\tau$ is fixed by the compactness parameter
(i.e. by the luminosity for fixed geometry). This yields a definite
relation between spectral variations and intensity, predicting modest
spectral changes for large (factor 10 at least) variations in the
intensity. On the contrary, for low pair density plasmas, significant
spectral variations are possible even in the absence of luminosity
variations.  { In the latter case, for constant geometry (i.e. at a
constant Compton parameter $y$), the spectral index is expected to
increase when the electron temperature decreases (Haardt, Maraschi \&
Ghisellini \cite{haa97})}.

In the present paper, our aim is to test Comptonization models over high
quality data, deriving { further} constraints on the physical parameters
and geometry of the source.  To achieve such goal, the long look at the
Seyfert I galaxy NGC 5548 performed by \BS provides an ideal
dataset. Previous studies of the source conducted over several years
using EXOSAT, GINGA, ASCA and RXTE and, in some cases, coordinated
observations first with IUE and later with EUVE (Walter \& Courvoisier,
\cite{wal90}; Nandra et al., \cite{nan91}; Chiang et al., \cite{chi00})
support the general Comptonization picture, revealing the presence of
several components in the X-ray spectra (neutral Iron line, reflection
hump, soft excess...). No information on the high energy and of the
Comptonization component could however be obtained from the above
studies. Magdziarz et al. (\cite{mag98}), using average OSSE data and non
simultaneaous GINGA observations, suggested a temperature of 50~keV for
the hot corona.

 NGC 5548 was observed by \BS in a single long (8 days) observation, with
a net exposure of 314 ks on source.  The high quality of the \BS data
allows a detailed study of the spectrum over a very wide energy range,
from 0.2 to 200 keV, and offers the opportunity to study spectral
variability, since a conspicuous flare occurred in the middle of the
observation. A detailed analysis of these data has been presented by
Nicastro et al. (\cite{nic99}, hereafter N99). The differences between
the latter analysis and our results concern only the modelling of the
continuum and will be discussed in the course of the paper.

Our main progress here is to adopt and fit directly to the data a
detailed model of the Comptonized spectrum, for which the commonly
adopted representation of a simple power law with a high energy cut--off
turns out to be a rather poor approximation. Effectively, Comptonized
spectra intrinsically show additional features, such as bumps due to
different scattering orders, and an anisotropy "break" due to the
(plausible) anisotropic nature of the soft photon input.

The paper is organized as follows. In section \ref{comptonmodel} we
briefly summarize the main characteristics of Comptonization models, and
compare results of different approaches and geometries. The analysis of
the \BS data is presented in section \ref{dataanal}.  We will not be
concerned here in detail with the warm absorber features already
discussed in N99. In section \ref{iueosse}, we compare the \BS data with
non--simultaneous IUE and OSSE data. We discuss our results and their
physical interpretations in section \ref{discussion}. We then conclude in
the last section.

\section{The Comptonization Model}
\label{comptonmodel}
Theoretical Comptonized spectra, produced by a mildly relativistic plasma
scattering off low--frequency radiation, have been computed for two
decades now (Shapiro, Lightman \& Eardley \cite{sha76}; Sunyaev \&
Titarchuk \cite{sun80}; Pozdnyakov, Sobol \& Sunyaev \cite{poz76}). More
recently, it has been realized that, if the scattering occurs above an
accretion disk, the source of seed photons is anisotropic, introducing
anisotropies and modifications of the outgoing spectrum. Haardt \&
Maraschi (\cite{haa91}, \cite{haa93b}) and Haardt (\cite{haa93a}) derived
the angle--dependent spectra from disk--corona system using an iterative
scattering method, where the scattering anisotropy was taken into account
only in the first scattering order.  Other more detailed works have
followed, exploiting non linear Monte Carlo techniques (Stern et al
\cite{ste95b}), or the iterative scattering method (Poutanen \& Svensson
\cite{pou96}, hereafter PS96), allowing the treatment of systems with
different geometries such as slabs, cylinders, hemispheres, and
spheres. All these works have shown that anisotropic effects are
important, and indeed can modify substantially the spectral shape of the
Comptonized radiation. The largest effect occurs when soft photons are
emitted by a plane (disk) on one side of the corona. In this case,
photons backscattered towards the disk in the first scattering are
necessarily produced in "head--on collisions" and therefore have an
energy gain larger than average, while photons scattered towards the
corona (i.e. in the forward direction) have an energy gain smaller than
average.  As a consequence, the contribution of the first scattering
order to the outgoing flux is significantly reduced. Clearly, this effect
becomes important when the energy gain per scattering is large, that is
when $kT_e$ is mildly relativistic, say $kT_e \gta 100$ keV.

In Fig. \ref{anisbreak} we compare spectra computed for the same value of
the coronal temperature ($kT_e$=360~keV) for different geometries.  For
the slab geometry, we show spectra obtained with the code of Haardt
(\cite{haa94}, hereafter H94), whereas the hemispherical and the
spherical ones have been produced by the code kindly made available by
Poutanen \& Svensson (PS96). We have checked that in the slab case the
two codes give identical results. For the sphere, the soft photons are
supposed to be emitted isotropically at the centre of the sphere, whereas
they come from the bottom for the slab and the hemisphere
configurations. In each case, the optical depths have been chosen so as
to produce approximatively the same spectral index in the X--ray
range. We have thus taken $\tau$=0.09, 0.16 and 0.33 for the slab,
hemisphere and sphere geometry, respectively.  We have also plotted, for
comparison, a cut--off power law spectrum setting the e--folding energy
$E_{\rm c}=2kT_{\rm e}$=720~keV, as a first order approximation to
Comptonization spectral models (for $\tau \lta 1$).

It is clear, from Fig. \ref{anisbreak}, that the spectra are quite
different at medium - high energy ($E \gta 10$ keV).  Below the high
energy cut--off the spectra for the slab and hemisphere cases can be
approximately described by broken power laws. The slope observed at low
energies is harder due to the deficiency of once scattered photons.  In
this range the spectral index depends on the viewing angle, since the
anisotropy is maximal at 0 degrees. The slope {\it above the break}
corresponds to that produced by the Comptonization process without taking
into account anisotropy and is therefore almost angle--independent.

The energy of the break $E_{\rm break}$ should approximately correspond
to the peak energy of the spectrum formed by twice--scattered photons.
In practice { $E_{\rm break}$ is close to the peak energy of the third
scattering order (Haardt 1993)}, that is (assuming a black body soft
emission):
\begin{equation}
E_{\rm break}\simeq 2.7kT_{\rm
bb}\left[16\left(\frac{kT_e}{m_ec^2}\right)^2+4\left(\frac{kT_e}{m_ec^2}
\right)+1\right]^3.
\label{eqbreak}
\end{equation}

In cases relevant for Seyfert galaxies, $E_{\rm break}$ is typically of
the order of few keV, much lower than the high energy cut--off.  The
anisotropy break is thus an important feature when fitting real data,
even if it may be hidden by the presence of the reflection hump above 10
keV.\\

The anisotropy break is important only for high temperatures.  In fact,
for $kT_e \lta 50$~keV, the forward--backward anisotropy becomes small
and consequently, the amplitude of the spectral break decreases. Also,
for small values of $\tau$, a broken power law is not a good description
of the emerging spectrum anyway, since the spectrum is formed by distinct
humps corresponding to the different Compton scattering orders.  Finally,
for high values of $kT_{\rm bb}$, as may be the case for galactic black
holes, $E_{\rm break}$ can be well above 10 keV, completely hidden by the
reflection component.

For the same temperature, a hemispheric geometry gives rise to a larger
anisotropy (larger break) than a slab. This can be qualitatively
understood as follows. Consider a pole on observer and the seed photons
emitted at { the centre of the bottom of the hemisphere}. In the first
scattering order all scattering angles between 0 and 90 degrees (final
direction towards the pole on observer) have similar probabilities.  In
the slab case the scattering at 90 degrees has larger probability
relative to that at 0 degrees since the optical depth along the
horizontal direction is larger than along the vertical.

Note that the shape of the break has important effects on the best--fit
values of other parameters involved in the fitting procedure. For
instance, we expect that the amplitude of the reflection component
derived { from the spectral fit} when allowing for an anisotropy break in
the continuum will be larger for configurations with larger anisotropy
and minimal in the case of a simple power-law model (cf. Table\ref{tab}).

\section{Application of different models to the \BS data of NGC 5548} 
\label{dataanal}
Here we will be concerned with data from three instruments on board \BS:
the Low Energy Concentrator Spectrometer, LECS (Parmar et al.
\cite{par97}) covering the 0.15--10 keV range, the Medium Energy
Concentrator Spectrometer, MECS (Boella et al. \cite{boe97}) covering the
2--10 keV range and the Phoswich detector system, PDS (Frontera et
al. \cite{fro97}) covering the range 12 -- 200 keV. For a description of
the instruments and data reduction procedure we refer to N99 and
references therein.
 
A careful analysis of the \BS data was carried out by N99. The continuum,
 was described using a simple cut--off power law model
\begin{equation}
f_{E} \propto E^{1-\Gamma}\, {\rm e}^{-E/E_{\rm c}},
\end{equation}
characterized by the photon index $\Gamma$, and the e--folding energy
$E_{\rm c}$, with additional, superimposed features due to reflection and
the possible presence of a warm absorber. The source brightened during
the central $\sim$70~ks of the observation by $\sim$30\% in the LECS and
by $\sim$15\% in the MECS. No significant variability larger than $\sim$
20\% was detected in the 13--200 keV PDS range. N99 divide the
observations into three parts, each with nearly constant intensity and
with similar signal to noise ratio: two low states L1, L2 (from the first
$\sim$ 120~ks and the last $\sim$ 106~ks of the observation respectively)
and a high state H (from the central $\sim$~70 ks of the
observation). The main results of N99 can be summarized as follows:

\begin{itemize}
\item the average spectral data can be well fitted by a cut--off power
law, plus reflection component and warm absorber features. For most of
the duration of the observation (i.e. in the low state), the continuum is
characterized by $\langle\Gamma\rangle =1.59$ and $E_{\rm c}=115$ keV

\item NGC 5548 shows significant spectral variability between the low and
high states, with $\Delta\Gamma\simeq$~0.2 (the low state being the
harder one), whereas the total luminosity remains essentially
constant. The high energy cut-off is also tightly constrained in the low
state, $E_{\rm c}=115_{-27}^{+39}$~keV, while, in the high state, a lower
limit is obtained, $E_{\rm c}\gta 260$ keV. This lower limit is however
larger than the one obtained in the low state.

\item the presence of a warm absorber in this source is confirmed, and
for the first time an emission feature, possibly associated with the
OVII-OVIII K$\alpha$ and K$_{\beta}$ emission lines at $\sim$0.6 keV, is
detected.
\end{itemize}

Here we focus on the continuum plus reflection components fitting
directly Comptonization spectra computed for different values of the
physical parameters $kT_{bb}$, $\tau$ and $kT_e$ and for different
geometries.  In our analysis, we used the H94 and PS96 codes.  The
advantage of the H94 code is that its outputs are computed before the
fitting procedure, and stored as a table in {\sc xspec}, so that
searching for best fit values, confidence levels, errors, etc, is a very
fast procedure.  For this reason, throughout the paper, fits to the slab
model always refer to the H94 code outputs, while for the hemisphere and
sphere geometries the results reported have been obtained using the PS96
code.

The reflection component was included following White, Lightman \&
Zdziarski (\cite{whi88}) and Lightman \& White (\cite{lig88}) and
assuming neutral matter. In the H94 code, a constant spectral shape
averaged over angles is assumed for the reflected photons. {It is
multiplied by a normalization factor which depends on the inclination
angle} (see Ghisellini, Haardt \& Matt 1994 for details). PS96 instead
computes the reflection component for the given inclination angle using
the Green functions of Magdziarz \& Zdziarski (\cite{mag95}).  The
geometrical normalization of the reflection component $R$ is left as a
free parameter. A value $R=1$ corresponds to a covering factor of the
cold matter to the X--ray source of $\Omega=2\pi$. Note that the X--ray
emission, as discussed in Section \ref{comptonmodel}, is not assumed to
be isotropic.

The iron line is modeled by a Gaussian with null width, since the line
appeared practically unresolved ($\sigma_{F_e}<$ 300 eV) in the previous
analysis of N99. The central energy of the line is fixed at 6.3 keV, the
best fit value obtained by N99. For what concerns the warm absorber (WA),
we simply modelled it as two edges at 0.74 and 0.87~keV, to account for O
VII and O VIII absorption.  A more detailed analysis, as that done by
N99, is beyond the scope of our paper, which is instead focused on the
high energy continuum. We have however checked that a better modelization
of the WA (CLOUDY code used by N99) does not affect the results obtained
in this paper.

Finally, the LECS/MECS and PDS/MECS normalization ratios were frozen to
the standard values of 0.75 and 0.88 respectively (Fiore, Guainazzi \&
Grandi \cite{fio99}). The results of the fits are detailed in the
following sections. Throughout the paper, errors on single parameter
(obtained by letting the others free to vary) are quoted at a confidence
level of 90 \% (i.e. $\Delta\chi^2$=2.7) unless otherwise specified.\\


\subsection{Constraining the soft photon field and the inclination angle}
\label{dataanala}
To take advantage of the maximum available signal to noise ratio, we
consider first the total data set obtained by \BS, that is the average
over the whole observation period from the three Narrow Field
Instruments, LECS, MECS and PDS. These data taken together cover the
range from 0.15 to 200 keV.

We start to constrain the values of the soft photon temperature $kT_{\rm
bb}$, and of the inclination angle $i$ (or its cosine $\mu \equiv \cos
i$). To this end we assume a fixed, slab geometry where the input soft
radiation field is supposed to be localized on one side of the slab.

A soft X--ray excess was found previously in NGC 5548 (Done et al.
\cite{don95}; Magdziarz et al. \cite{mag98}), but is not present in the
\BS spectra (N99).  From IUE data, the UV bump seems to peak above 3 eV
(Clavel et al. \cite{cla92}; Magdziarz et al. \cite{mag98}). We therefore
tested values of $kT_{\rm bb}$, in the 5 to 50 eV range.
  
We perform a best--fit analysis on the four parameters $kT_{\rm bb}$ ,
$kT_e$, $\tau$ and $R$ with two fixed values of $\mu$. The results of the
fitting procedure are reported in Table 1, where we give the best fit
values and 90\% confidence errors for each of the four fitted quantities.

For an inclination angle of 30$^{\circ}$ the best fitting soft
temperature $kT_{\rm bb}$ is 16 eV, with a 90\% confidence interval
between 7 and 18 eV.  This is higher but not far from the maximum
temperature deduced from the UV spectrum suggesting that the emission
from the accretion disk should extend to higher energies. This could be
due to deviations from a pure blackbody emission and/or to the presence
of a hotter inner region of the disk.  The reflection component is nicely
consistent with a solid angle of 2$\pi$.  High black body temperatures
(20 eV and larger) are statiscally unacceptable. The reason is that, in
these cases, the high energy tail of the UV--to--Soft--X-ray bump would
fall in the LECS band.

It is worth stressing, at this point, that the \BS data, as already noted
by N99, do not confirm the presence of a soft excess, as observed by
ROSAT in 1992-1993 (Done et al. \cite{don95}). This discrepancy could be
due to systematic errors in the ROSAT low energy data, as already
discussed by Iwasawa et al. (\cite{iwa99}) but may also result from a
possible variability of the soft component. In the latter case the soft
photon temperature could be different at different epochs. We recall that
the soft excess derived by Reynolds (\cite{rey97}), fitting ASCA data of
NGC~5548, having best fit black body temperature of 240 eV, referred to a
state when the spectral index of the continuum was 1.9, definitely
steeper than during this {\it Beppo}SAX observation (see further
discussion section \ref{sectvaria}).

For a larger inclination angle, $i=60^{\circ}$, the fit yields a smaller
value of the reflection component $R$.  This is because the anisotropy of
the Compton emission is maximum at $i=0^{\circ}$, and is reduced for
increasing inclination.  Hence the difference in spectral index before
and after $E_{\rm break}$ is smaller for i$=60^{\circ}$. The MECS data
constrain the spectral index for $E \lta E_{\rm break}$, leaving, for
larger viewing angles, less room for the reflection component, which
peaks at energies $E\gta E_{\rm break}$.  Figure \ref{bestfitslabhemi}
shows the LECS, MECS and PDS spectra and the best fit models obtained
with inclination angles of 30$^{\circ}$ and 60$^{\circ}$ respectively.
Although the $\chi^2$ value is acceptable in both cases, it is
significantly worse for the larger angle ($\Delta\chi^2$=35) as shown by
the fit residuals in the PDS range.  We will not discuss the inclination
further fixing $i=30^{\circ}$ in the following.

The $\chi^2$ contour plots for two parameters (confidence levels of 68\%,
90\%, and 99\%) for slab models with $i=30^{\circ}$ are shown in
Fig. \ref{contplot}. The joint confidence levels of $kT_e$ and $\tau$ is
shown in Fig. \ref{contplot}c. To compute these contour plots $kT_{bb}$,
$\tau$, $kT_e$, $R$ and the model normalization were allowed to vary,
while the Fe line intensity and the depth of the Oxygen edges were frozen
to their best fit values, { since they do not significantly affect the
fit of the continuum}. From such plot one can see how $kT_e$ and $\tau$
adjust so that the combination of the two produces a spectrum with the
observed shape. In fact the shape of the contours follows approximately a
line of constant Comptonization parameter.

\subsection{Comparing different geometries}
\label{compdiffgeom}
In order to test different geometries we performed spectral fits using
the PS96 code for a hemisphere with soft photon input from the bottom
plane and a sphere with soft photons isotropically emitted at the centre.
In the first case the fits were performed for different values of the
inclination angle as previously done for the slab case. Again the best
quality fits are obtained for $i=30^{\circ}$. The results are reported in
Table \ref{tab}. In both cases the derived temperature of the hot
Comptonizing electrons does not change significantly with respect to the
case of the slab geometry.  The optical depth adjusts as expected towards
somewhat higher values, since for the slab configuration the optical
depth parameter, which is measured in the vertical direction, is lower
than the average optical depth seen by the Comptonized photons. The soft
photon temperature for the hemisphere model also coincides with that for
the slab, while it is somewhat lower in the case of the sphere. The
largest difference is the value of the reflection component, which is
larger for larger anisotropy.  In fact the hemisphere yields the largest
value and the sphere the smallest one (cf. section \ref{comptonmodel}).

We also report in Table \ref{tab} the best fit parameters of a simple
cut--off power law (plus reflection) model ({\sc pexrav} model in {\sc
xspec}, Magdziarz \& Zdziarski \cite{mag95}). For this model, the fit
parameters are the e--folding energy of the cut--off power law $E_{\rm
c}$, the photon index $\Gamma$ and the reflection normalization $R$.
Note that our best fit values are in complete agreement with those
obtained by N99. On the other hand, here and in the following sections,
we report ``equivalent" values of $kT_e$ also for {\sc pexrav}, simply
computed as $kT_e\equiv E_{\rm c}/2$, keeping in mind that such
approximation roughly holds for $\tau\lta 1$. For $\tau \gg 1$,
$kT_e\equiv E_{\rm c}/3$ would be more appropriate. The value reported
can be considered as an upper limit to the temperature estimated from a
cut--off power law model. Knowing the temperature, the spectral index
derived from the {\sc pexrav} fit can be used to determine the optical
depth using the relation (Shapiro, Lightman \& Eardley \cite{sha76};
Sunyaev \& Titarchuk \cite{sun80}; Lightman \& Zdziarski \cite{lig87}):
\begin{equation}
\Gamma-1=\displaystyle\left[\frac{9}{4}+\frac{m_ec^2}{kT_e
\tau(1+\tau/3)}\right]^{1/2}-\frac{3}{2}
\end{equation}
This equation is valid for $\tau \gg 1$, and we have checked {\it a
posteriori} that such condition is roughly matched in all cases.

Every alternative geometry gives statistically acceptable fits. The
spherical (isotropic) model, though acceptable, is however considerably
worse than the others. Also it is surprizing that the best fit
temperature obtained with this model is similar to the anisotropic models
rather than to the {\sc pexrav} results. We therefore checked this result
fitting the data with the analytical Comptonization model of Titarchuck
({\sc comptt} model of {\sc xspec}, Titarchuk \cite{tit94}) and adding
the reflection component computed by {\sc pexrav}, a gaussian for the
iron line and two edges for the O VII and OVIII ones. Such model should
represent quite well the case of an isotropic spherical geometry.  The
best fit then gives $\tau\simeq 4.3$ and $kT_e\simeq$ 30 keV, comparable
to the {\sc pexrav} results if $kT_e\simeq E_c/3$.  We believe that the
discrepancy with the results obtained using the spherical code of
Poutanen is due to the fact that the latter is limited to $\tau\leq 3$ {
(we have cheked that both codes are in agreement when we fixed $\tau$ to
3)}. This can prevent the fitting procedure to find an acceptable fit
with low temperature and high optical depth. We therefore will not
discuss the spherical model further.

The data and best fit model for the hemisphere are compared in
Fig. \ref{bestfitslabhemi} with those for the slab.  In
Fig. \ref{contplotb}, we show the confidence levels for various
combinations of the fitted parameters (here again only $kT_{bb}$, $\tau$,
$kT_e$, $R$ are allowed to vary) for the hemisphere configuration for an
inclination angle fixed at $30^{\circ}$. In particular the contour plots
of $kT_e$ and $\tau$ for the hemisphere do not overlap those for the
slab. Although the temperature range is the same, the optical depth is
higher, yielding a larger value of the Comptonization parameter.

The theoretical relation between $kT_e$ and $\tau$ expected for the slab
and hemisphere configurations in energy balance and with negligible disk
heating as computed by Stern et al. (\cite{ste95a}) are also shown on the
same figures. The parameters derived for the slab correspond to a
Comptonization parameter $y\simeq$ 1 higher than expected in an
equilibrium configuration ($y\simeq 0.5$). The case of a hemisphere
appears in better agreement with the theoretical predictions from energy
balance requirements. For the latter model the normalization of the
reflection component is however relatively large (R $=1.6\pm 0.3$) in {
marginal agreement} with the measured EW (of the order of 120 eV) of the
Fe line (George \& Fabian \cite{geo91}; Matt et al. \cite{mat91};
Reynolds, Fabian \& Inoue \cite{rey95}). Thus neither model is entirely
satisfactory.  This problem will be further discussed in section
\ref{discussion}.

The most important and general conclusion from this section is that the
temperature estimates of the hot electrons derived from fits with
Comptonization spectra, irrespective of geometry, are substantially
different from those obtained using models involving a power law with
high energy cut--off plus a reflection component, exemplified by {\sc
pexrav} fits (see Table 1 and 2). This is due to the fact that the power
law slope is determined with small errors by the LECS and MECS
data. Since within {\sc pexrav} type models this slope by hypothesis
cannot change at higher energies, a cut--off around 100 keV is required
to fit the PDS data. In Comptonization models, the LECS and MECS data
determine the slope below the anisotropy break. Above this break the
intrinsic spectrum is steeper and thus can fit the PDS data without an
additional steepening below 100 keV, allowing for a larger temperature.
The large difference in the fitted temperatures leads to widely different
predictions as to the fluxes emitted at higher energies, above the PDS
range.

\subsection{Comparing the low and high states}
\label{lhstate}

We now discuss the spectral variations associated with the flare
experienced by NGC~5548 during the central $\sim$70 ks of the \BS
observation.  Using direct Comptonization fits we can derive the
variations of the physical parameters during the flare. Since no
significant variations were found by N99 between the two low state
spectra (before and after the flare), we combined them into a mean low
state spectrum, labelled as ``L". The high state spectrum will be
referred to as ``H". Again, we fixed the inclination at 30$^{\circ}$ and
performed the fitting procedure for the plane parallel and hemispherical
geometries for the two states.

The fits to the L and H spectra are shown in Fig. \ref{speclemepdls}.
The high quality of the data allows to constrain quite precisely the main
parameters $kT_{bb}$, $\tau$, $kT_e$ and $R$ in the two states. The best
fit values are reported in Table \ref{tab2}, with their associated 90\%
error and confidence level contours are shown in Fig
\ref{contplotlh}. While these errors take into account the possible
variations of all the parameters involved, we stress that not all the
parameters are expected to vary simultaneously. To illustrate the mutual
parameter dependence we also show in Fig \ref{contplotlh}c the contour
plots obtained assuming that the soft photon temperature remained
constant during the flare.  These "partial" contours suggest that for
constant $kT_{\rm bb}$ the change in physical parameters is essentially a
decrease of the coronal temperature. This is even more true if $kT_{\rm
bb}$ {\it increases} in the softer state as indicated by the best fit
values and discussed in section \ref{sectvaria}. We have reported the
values of $kT_e$ and $\tau$ found for the low and high states for the
slab and hemisphere configuration in Fig. \ref{diffgeo}. Two main points
should be noted here:

\begin{itemize}
\item the change of state of the source, from L to H, is consistent with
a constant Compton parameter $y$. The trend in the $\tau$ -- $kT_e$ plane
is for a decrease of the temperature of the hot corona in the high state,
with a roughly constant optical depth.

\item As the high state is the softer one (cf. Table \ref{tab2}), the
temperature $kT_e$ and the slope of the intrinsic continuum are
anticorrelated, as expected if an increase of the cooling caused the
hard--to--soft luminosity ratio to decrease.
\end{itemize}

We also derived values of $kT_e$ and $\tau$ for both the low and high
states using the slopes and cut--off energies obtained from fits with
{\sc pexrav} (see section 3.2) . These are reported in Table \ref{tab2}
and shown in Fig. \ref{diffgeo}.  The errors on the temperature are
rather large, since in this kind of model it is derived from the cut off
energy which is constrained only by the hard X-ray data. { We recall that
in N99, only a lower limit to the cut--off energy was obtained for the
high state. Our results have somewhat tighter constraints because we have
fixed the relative normalization of the instruments.}

In contrast with the results obtained using anisotropic Comptonization
the parameter variations derived fitting {\sc pexrav} models suggest
that:
\begin{itemize}
\item the change of state of the source is consistent with the Compton
parameter remaining constant during the transition but the trend in the
$\tau$ -- $kT_e$ plane is for an increase of the temperature of the hot
corona in the high state, with a decrease of the optical depth.

\item the temperature of the corona is thus positively correlated with
the continuum spectral index.
\end{itemize}

Thus not only different continuum models yield different values of the
physical parameters but also the variability trends derived are opposite.

\section{Multifrequency SED using non simultaneous IUE and OSSE data}
\label{iueosse}
 Our purpose here is to examine whether existing data on the broad band
energy distribution (optical to hard X--ray) may provide additional
constraints to the Comptonization models discussed above.  We therefore
compare the \BS data on NGC 5548 to non--simultaneous UV and soft
$\gamma$--ray data, obtained with IUE, and with OSSE on board {\it CGRO},
respectively. These data have been analysed, together with Ginga and
ROSAT data, in Magdziarz et al. (\cite{mag98}, hereafter M98).

\subsection{{\it UV data}}
\label{iue}
The ultraviolet spectra were extracted from the IUE electronic archive,
and cover nine epochs, from January 1989 to July 1990.  We use the data
as reduced by M98.  Emission lines were removed, and dereddened with
E(B-V)$\simeq$ 0.03.
The fit of the IUE data shown in Fig. \ref{n5548spec} was done with an
accretion disk model, consisting of multiple blackbody components,
modified by the estimated absorption. In addition a vertical line
represents the dispersion of the UV fluxes of NGC 5548 measured with IUE
between 1983--1995 (see also Clavel et al. \cite{cla92}). The best fit
gives an inner disk temperature of $\sim$ 3 eV and a flux of $\sim
10^{-10}$~erg~s$^{-1}$~cm$^{-2}$. Note that the UV and X--ray
luminosities appear to be of the same order.  Moreover the extrapolation
of the X-ray spectrum towards lower frequencies falls well below the peak
of the putative disk emission.  Although some caution is in order due to
the non simultaneity of the observations and to the absence of data in
the FUV -- EUV range, this result is consistent with a) a corona with low
optical depth and covering factor of order unity or b) a patchy corona
with larger ($>$ 1) local optical depth but which covers a small fraction
of the surface which emits the thermal radiation.

\subsection{{\it OSSE data}}
\label{osse}
The OSSE data were obtained during different observing periods in 1991
and in 1993 ( See also M98). Their average was used to constrain the high
energy cut-off by fitting simultaneously the \BS data with the OSSE data
with a free relative normalization, which turns out to be 30\% higher
with respect to the MECS fluxes. Unfortunately, the OSSE data have a
relatively poor signal--to--noise ratio, and their inclusion in the fit
does not lead to an improvement in the determination of the temperature
of the hot plasma. They clearly show however that the larger temperature
estimated with the Comptonization models is completely consistent with
the highest energy data available.

\section{Discussion}
\label{discussion}

\subsection{Physical parameters of the corona}
\subsubsection{Temperature and optical depth}
Thanks to the long \BS observation of NGC 5548, we were able to test
quantitatively accurate anisotropic Comptonization models for the X-ray
emission from this source.  We have shown that if such models are
adopted, either in a plane parallel or hemispheric approximation for the
geometry of the corona, the inclination angle should be low, $i \simeq
30^{\circ}$ and the soft photon temperature $kT_{\rm bb} \simeq$ 5-15 eV.

The best fit values of the hot plasma temperature and optical depth
depend somewhat on the adopted geometry and on the state of the source,
but in any case, {\it the estimated temperatures, $kT_{\rm e}\simeq$ 250
-- 260 keV, are substantially larger than those inferred from fits with a
cut--off power law}. N99 find a high energy cut--off $E_{\rm c}$ of the
order of 120 keV, corresponding to a temperature $kT_{\rm e} \simeq 60$
keV (see also Table \ref{tab2}).

Correspondingly the optical depth, $\tau\simeq$ 0.16--0.37 (for the slab
and hemisphere geometry) is smaller than that deduced from the {\sc
pexrav} best fit where $\tau\simeq$ 2.4. Large values of $\tau$ should
have observable consequences on the profile of the Fe line component
produced in the disk, leading to an attenuation and broadening of the
line. At present the profile of this line in NGC 5548 is not
unambiguously determined.  The line seems unresolved in the \BS data
(N99) but a significant width is suggested by ASCA data (Mushotzky et
al., \cite{mus95}; Nandra et al., \cite{nan97}; Chiang et al.,
\cite{chi00}; N99). Forthcoming observations with CHANDRA and XMM-Newton
are expected to bring substantial progress on this issue.

We show in Fig.\ref{n5548spec} the best fit Comptonization model (in slab
geometry) of the \BS data of NGC 5548, together with the best fit model
obtained with a cut--off power law model (in solid and dashed line
respectively). Due to the different normalization of the reflection
component both models are in good agreement for $E\lta 100$ keV. Above
this energy the two models diverge, the first one predicting much larger
fluxes. Unfortunately the available non--simultaneous OSSE data cannot
discriminate between the two. This underlines the need of better soft
$\gamma$--ray observations, as those possibly provided by INTEGRAL, in
order to directly confirm or disprove the temperatures inferred with
anisotropic Comptonization models.

\subsubsection{Energy Balance}
The physical parameters $kT_{\rm e}$ and $\tau$ derived for the corona in
the case of a plane parallel geometry are not in agreement with the
predictions for a simple two phase disk--corona system in energy balance,
even under the extreme assumption that all the available gravitational
power is dissipated in the hot corona (Haardt \& Maraschi \cite{haa91},
\cite{haa93b}; Stern et al. \cite{ste95a}). They instead suggest that the
hot gas is {\it photon starved}, i.e., it is undercooled with respect to
theoretical expectations based on a stationary equilibrium hypothesis.

Indeed, the fits obtained with a hemispherical model
(cf. Fig. \ref{contplotb}) yield parameters in agreement with theoretical
predictions for this geometry under the same assumptions mentioned above.
In fact, a hemisphere is naturally photon starved with respect to an
(infinite) plane parallel system, because of ``side effects'', which are
obviously negligible when the scale height of the corona is much lower
than its radial dimension.  In any case the best fit values of $kT_e$
are, in the two geometries, roughly consistent, while there is a factor
of 2 of difference between the corresponding values of $\tau$ (on this
point, see section \ref{comptonmodel}).

In the hemisphere case, however, we obtain a large normalization of the
reflection component, in excess of unity (again, see section
\ref{comptonmodel} and Fig. \ref{anisbreak} on this point), which does
not agree with the observed EW of the Fe line, while the slab geometry
gives a slightly better fit with a reflection of order unity, as
expected. These results indicate that the real model is probably more
complicated but not too far from these two ideal cases.

Our results indicate that the anisotropy break should be as expected for
the slab but the Comptonization parameter should be higher, as expected
for the hemisphere. These two aspects could be reconciled if a
significant fraction of the area below a flattish corona {\it did not}
contribute to absorb and reprocess the downscattered Comptonized photons
into soft luminosity. One could envisage a picture in which the corona
extends down to the very innermost orbits around the black hole, while
the inner part of the disk, could be hot and highly ionised. In this kind
of picture the average energy balance is shifted toward a photon starved
condition, while most of the reflection features from the external part
of the disk should be visible anyway. Detailed calculations are necessary
to test such kind of picture. A model along this line has been proposed
by Belloni et al.  (\cite{bel97}) for GRS 1915+105, and is believed to be
relevant in the context of the different spectral states observed in
Galactic black holes (see, e.g., Narayan, Mahadevan \& Quataert
\cite{nar98}).

\subsection{Variability}
\label{sectvaria}
As already discussed, the source exhibits a flare in the central part of
the observation during which the X-ray spectrum steepens. Analysing the
spectral variability in terms of anisotropic Comptonization models, we
find that, irrespective of geometry (see section \ref{lhstate}, and
figure \ref{contplotlh}), the best fit parameters for the low--to--high
state suggest a change of the Compton parameter y, in the sense that y
tends to decrease in the high state (cf. Fig. \ref{diffgeo}). A decrease
of $y$ implies a decrease of the Comptonized to soft luminosity ratio.

Since the X--ray continuum varies more at lower energies, pivoting at the
highest energies, where the bulk of the Comptonized luminosity is
emitted, the change in $y$ is most likely due to an increase of the
cooling, rather than to a decrease of the heating. In fact if the heating
decreased while the cooling remained constant the pivot should occur at
the lowest energies.  If this interpretation is correct, then the
spectral softening in the high state is naturally explained by a drop of
the coronal temperature, ultimately due to an increase of the EUV soft
photon flux. In fact our fits yield lower temperatures { of the corona}
in the higher state and are consistent with the optical depth remaining
constant in the transition. More precisely, from the high and low state
fits we estimate a Compton parameter of $\simeq 1.3 $ and $\simeq 0.6 $
in the low and high states respectively, corresponding within the model
to an increase of the soft photon luminosity of a factor $\simeq 3 $. {
It is worth noting that similar variability amplitudes have been observed
by EUVE in this source on a time scale of a few tenths of kiloseconds,
while the 2-10 keV flux was varying by about 20\% (Chiang et al.,
\cite{chi00}).}

We show in Fig. \ref{speclemepdlsmo} the full best fit models of the low
and high state including the soft photon component but removing the
neutral absorption and Oxygen edges for clarity. It is remarkable that
the required soft luminosities compare well with the intensity and
variability range of the UV spectra observed with IUE and shown in
Fig. \ref{n5548spec}. They require however, especially in the case of the
softer state, regions of higher temperature than visible in the UV range.
It is interesting to recall that Reynolds (\cite{rey97}), fitting ASCA
data of NGC~5548 found a significant soft excess with black body
temperature near 240 eV and a spectral index of the continuum of 1.9 in
the 2--10 keV range, definitely steeper than during this {\it Beppo}SAX
observation.  The above soft component could be the signature of the
UV-EUV seed photon becoming prominent enough to be directly detected. The
corresponding strong cooling of the corona would then imply a steeper
X-ray continuum in the ASCA data as observed.

Qualitatively, the spectral and luminosity change can be explained
assuming that, in the H (soft) state, the disk pushes a bit closer to the
black hole, increasing the portion of the hot corona which "sees" a
copious flux of UV--EUV soft radiation. This would cause a more effective
cooling of the hot gas, a decrease in its temperature, while the overall
energetics would not change substantially. A clear prediction is that the
temperature of the soft photons and the strength of the reflection
features should be larger in the H state with respect to the L state. Our
data are consistent with such a view, and, again, detailed calculations
are necessary to test the model quantitatively.

Interpreting variability in terms of the physical parameters obtained
with {\sc pexrav} leads to different conclusions. In this case the
Compton parameter is consistent with remaining constant between L and
H. The decrease of the optical depth is thus expected to go with an
increase of the temperature, in agreement with the best fit values. It is
however difficult to predict the corresponding change of the soft photon
flux.

\subsection{The role of pairs}
It is interesting to ask what could be the role of electron positron
pairs in a corona with the derived parameters. This is determined by the
"compactness" parameter $\ell_{\rm X}$.

For any given temperature, there exists a maximum value that $\ell_{\rm
X}$ can have, without violating pair equilibrium, i.e., pair production
balanced by pair annihilation (Haardt \& Maraschi \cite{haa93b}; Svensson
\cite{sve96}). Such limit compactness is reached in the case of a {\it
pure pair} plasma, a situation in which the pair production is so intense
that the new created particles dominates the scattering opacity against
"normal", pre--existing electrons. A further increase of $\ell_{\rm X}$
would produce more pairs than annihilate, increasing the net opacity of
the source, and hence the cooling. The plasma can find a new equilbrium
only at a lower temperature.

We can estimate the compactness of NGC 5548 from the luminosity ($L_{\rm
X} \simeq 10^{44}$ erg s$^{-1}$), and from the variability time
scale. The variation observed during the \BS long look gives an upper
limit of $\sim 3 \times 10^{15}$ cm to the linear dimensions of the
varying region. From these values we derive a compactness
\begin{equation}
\displaystyle \ell_{\rm X} \equiv \frac{L_{\rm X}\sigma_{\rm T}}{R m_{\rm e} 
c^3}\simeq 0.1.
\end{equation} 
Such value is below the maximum allowed for a slab or hemisphere corona
with a temperature $\sim 250$ keV, assuming energy balance (Stern et al.
\cite{ste95b}), which is of the order of the unity. For this compactness
and this temperature the optical depth associated with pairs in
equilibrium is smaller than that estimated from our model fits.  However,
our estimate of the source dimension is probably a { rough} upper limit,
which translates into a lower limit for the compactness. A higher
compactness allows for more pairs.

In view of these uncertainities, we can only conclude that a pair
dominated corona is not required for NGC 5548.

\section{Conclusions}
The aim of this paper was to test realistic Comptonization models over
the high signal to noise \BS observations of NGC 5548 (which has been
detailed by N99). Our main effort was to adopt detailed Comptonization
codes that treat carefully the Compton processes in different geometries,
especially for what concerns the possible anisotropy of the soft photon
field. This last point is crucial, since, in geometries like slabs or
hemispheres, the observed first order scattering humps is highly reduced
in comparison to the others, producing an anisotropy break just above the
averaged energy of twice scattered photons. Specifically, the \BS
observation of NGC 5548 has allowed us to show that:
\begin{itemize}
\item the data are well fitted by a plane parallel corona model with an
inclination angle of $30^{\circ}$ and a soft photon temperature between 5
and 15 eV. The corresponding best fit values of the hot plasma
temperature and optical depth are $kT_{\rm e}\simeq$ 250 keV and
$\tau\simeq$ 0.1 respectively. These values of $kT_{\rm e}$ and $\tau$
are however not in agreement with a radiatively balanced two phases
disk/corona system in plane parallel geometry, even under the assumption
that all the gravitationnal power is dissipated in the hot
corona. Instead the data suggest that the hot Comptonizing gas is
photon-starved. A better agreement is effectively obtained with a
hemispherical geometry, which is naturally undercooled with respect to
the slab model. However for the hemisphere the anisotropy break is
larger, yielding a large reflection component (R=1.6$\pm$0.3). This
suggests that the real configuration is probably more complex than these
two ideal cases.

\item the flare occurring during the central 70 ks part of the run
enabled us to interpret the spectral change of the source between
different states in terms of changes in the physical parameters of the
system.  Within the framework of anisotropic thermal Comptonisation the
change of state suggests a variation of the Compton parameter $y$, and
thus a change of the central configuration.  The spectral variability can
be most naturally explained by an increase of the soft photon flux
causing a reduction of the coronal temperature. 
be the first

\item the zero order cut--off power law approximation commonly used to
Comptonization spectra leads to quite different results. The
corresponding best fit values of the temperature and optical depth of the
corona are $kT_e\simeq$ 60 keV and $\tau\simeq$ 2.4. The bright state is
consistent with the Compton parameter remaining constant during the
spectral transition, but suggets an {\it increase of the coronal
temperature} and a {\it decrease of the optical depth}. This model is
statistically acceptable.

\item We expect that the substantial improvements in the spectral data at
low and medium X-ray energies achievable with CHANDRA and XMM-Newton
could discriminate between these alternative models if complemented by
simultaneous measurements at higher energies possibly provided by \BS.  A
decisive progress could come from improved soft $\gamma$--ray
observations like those expected in the near future by INTEGRAL,

\end{itemize}

\noindent
{\sl Acknowlegements:} We gratefully acknowledge J. Poutanen for
providing us his code and the anonymous referee for his helpful
comments. POP, FH and LM were supported by the European Commission under
contract number ERBFMRX-CT98-0195 ( TMR network "Accretion onto black
holes, compact stars and protostars") and by the Italian Ministry for
University and Research (MURST) under grant COFIN98-02-154100. GM and GCP
were supported by the Ministry for University and Research (MURST) under
grant COFIN98-02-32.


\clearpage

\vskip 1.5 truecm

\centerline{ \bf Figure Captions}

\vskip 1 truecm

\figcaption[]{Comptonized models for different geometries assuming
$kT_e=360$ keV and $kT_{bb}$=5 eV. The slab, hemisphere and sphere
geometries are plotted in solid, dashed and dot-dashed line
respectively. For the spherical geometry, the soft photons are supposed
to be emitted isotropically at the centre of the sphere, whereas they
come from the bottom plane for the slab and the hemisphere. The optical
depths, chosen so as to produce approximately the same spectral index for
$E\lta 10$ keV, are 0.09, 0.16 and 0.33 for the slab, hemisphere, and
sphere geometry, respectively.  For comparison, we have also over-plotted
a cut--off power law with $E_{\rm c}=2kT_{\rm e}$ in dotted
line. \label{anisbreak}}

\figcaption[]{LECS, MECS and PDS count rate best fit spectra of NGC 5548
obtained with a slab (for 2 inclination angles $i$=30$^{\circ}$ and
60$^{\circ}$) and a hemisphere (for $i$=30$^{\circ}$) configuration. The
corresponding best fit values of $kT_{bb}$, $R$, $\tau$ and $kT_e$ are
reported in Table \ref{tab}.{ The ratio between the data and the model is
shown in the bottom plot of each figure} \label{bestfitslabhemi}}

\figcaption[]{Contour plots (confidence levels of 68\%, 90\%, and 99\%)
of a) $\tau$ vs. $kT_{bb}$, b) $kT_e$ vs. $kT_{bb}$, c) $\tau$ vs. $kT_e$
and d) R versus $kT_e$ for the slab geometry. In Fig. \ref{contplot}c we
have over--plotted the expected relation in case of a slab Comptonizing
region in energy balance (solid rectangle, solid line, from Stern et
al. \cite{ste95a}) as well as curves of constant Compton parameter in
dot-dashed line. \label{contplot}}

\figcaption[]{Contour plots as in Fig.\ref{contplot} but for the
hemispherical configuration. In Fig. \ref{contplotb}c we have
over--plotted the expected relation in case of a hemispherical
Comptonizing region in energy balance (solid hemisphere, dashed line,
from Stern et al. \cite{ste95a}) as well as curves of constant Compton
parameter in dot-dashed line. \label{contplotb}}

\figcaption[]{{ LECS, MECS and PDS spectra of the low (upper panel) and
high (lower panel) states of NGC 5548 together with the best fitted
Comptonization model for a slab configuration. The corresponding best fit
values of $kT_{bb}$, $R$, $\tau$ and $kT_e$ are reported in Table
\ref{tab2}.  The ratio between the data and the model is shown in the
bottom plot of each figure}\label{speclemepdls}}

\figcaption[]{Confidence levels (68\%, 90\%, and 99\%) for two parameters
($kT_e$ and $\tau$) for the low and high state for the slab (top and
bottom left) and the hemisphere (top and bottom right) configuration. In
a) and b), the soft temperature $kT_{bb}$ was let free to vary (and the
best fit values are indicated by a +), whereas in c) and d) it is fixed
to the extreme values allowed by the fit (reported on the figures). As in
Fig. \ref{contplot} and \ref{contplotb}, we have also over--plotted, for
both geometries, the $kT_e$--$\tau$ relation predicted for a corona in
energy balance (from Stern et al. \cite{ste95a}).\label{contplotlh}}

\figcaption[]{Best fit values of $kT_e$ and $\tau$ derived with different
models. Triangles and circles represent the results for anisotropic
Comptonization models in the slab and hemisphere geometries
respectively. Squares represent the values of $kT_e$ and $\tau$ obtained
with {\sc pexrav} fits (see text for details).  Filled symbols refer to
the high state, empty symbols to the low state. We have over-plotted the
$kT_e-\tau$ relations predicted when energy balance is achieved (solid
line for slab, and dashed line for hemisphere; from Stern et
al. \cite{ste95a}). The axis are in logarithmic scales.\label{diffgeo}}

\figcaption[]{The full deconvolved \BS spectrum of NGC 5548 in the low
state together with non--simultaneous IUE and OSSE data. We have
over--plotted the corresponding best fits obtained with an anisotropic
Comptonization model (slab geometry) -- solid line -- and a simple
cut--off power law model -- dashed line.  The IUE data have been fit with
a multiple blackbody accretion disk (cf. section \ref{iue}). The vertical
line represents the dispersion of the UV fluxes of NGC 5548 measured with
IUE between 1983--1995. For clarity, SAX PDS data are plotted with filled
circles and the OSSE ones with triangles. The IUE and OSSE data are from
M98.\label{n5548spec}}

\figcaption[]{{ Intrinsic model spectra including the required soft 
blackbody component for the low (dashed line)
and high (solid line) states { obtained from spectral fits with  Comptonization
models (slab geometry). The corresponding best fit values of $kT_{bb}$, $R$, $\tau$ and
$kT_e$ are reported in Table \ref{tab2} }. Galactic and warm absorption
features have been removed}.\label{speclemepdlsmo}}

\clearpage

\begin{table}[h]
\begin{tabular}{lccccccc}
\tableline
\tableline
Geometry&Inclination&$kT_{bb}$(eV) & $kT_{e}$(keV) & $\tau$ & $\Gamma$ &R & 
$\chi^{2}/dof$  \\
\tableline
Slab&30$^{\circ}$ &16$_{-9}^{+2}$ & 260$_{-20}^{+50}$ &
0.16$_{-0.04}^{+0.02}$ & - & 0.9$\pm 0.2$ & 133/168\\
&60$^{\circ}$&25$_{-3}^{+2}$& 210$_{-20}^{+20}$& 0.33$_{-0.04}^{+0.07}$& - 
&0.6$\pm 0.2$ &168/168\\
\tableline
Hemisphere&30$^{\circ}$ 
&14$_{-8}^{+2}$&250$_{-20}^{+70}$&0.37$_{-0.11}^{+0.06}$& -
&1.6$\pm0.3$&139/168\\ 
\tableline
Sphere& - &8$_{-3}^{+4}$&280$_{-4}^{+16}$&0.44$_{-0.02}^{+0.05}$&-&0.2$\pm 
0.08$&177/168\\ 
\tableline
{\sc pexrav}& - & - &60$_{-10}^{+25}$& 2.4$_{-0.5}^{+0.3}$ & 
1.59$_{-0.02}^{+0.01}$ & 0.5$\pm0.2$&153/169\\
\tableline
\end{tabular}
\caption{Best fit values of $kT_{bb}$, $kT_e$, $\tau$ and the reflection
normalization $R$, for different values of the inclination angle ${\rm
i}={\rm acos}(\mu )$ and for different geometries. For comparison, we
also show the best fit values obtained with {\sc pexrav}. We assume in
this case $kT_e=E_{\rm c}/2$ and we compute $\tau$ as explained in
section \ref{compdiffgeom}.}
\label{tab}
\end{table}

\newpage

\vskip 2.5 truecm
\vspace*{2.5cm}

\begin{table}[h]
\begin{tabular}{llcccccc}
\tableline 
\tableline 
State&Geometry& $kT_{bb}$(eV)& $kT_e$(keV)&$\tau$  &$\Gamma$&R  &
$\chi^2/dof$ \\ 
\tableline 
Low&Slab&$8^{+10}_{-4}$&$330^{+70}_{-80}$&$0.12^{+0.08}_{-0.04}$
&-&0.9$\pm 0.2$  & 82/113\\
&Hemisphere&5$^{+12}_{-3}$&360$^{+80}_{-120}$&0.21$^{+0.28}_{-0.06}$&-&1.8$\pm 
0.3$&80/113\\ 
&{\sc pexrav}&-&$55^{+25}_{-10}$ 
&2.6$^{+0.2}_{-0.6}$&1.55$_{-0.02}^{+0.02}$&0.5$_{-0.2}^{+0.2}$&93/114\\ 
\tableline
High&Slab&$15_{-10}^{+2}$&$245^{+55}_{-30}$ &$0.12^{+0.04}_{-0.05}$ &-&1.0$\pm 
0.4$ &135/144\\ 
&Hemisphere&13$^{+2}_{-8}$&235$^{+65}_{-20}$&0.27$^{+0.08}_{-0.11}$&-&2.2$\pm 
0.5$&142/144\\ 
&{\sc pexrav}&-&$80^{+200}_{-35}$ 
&1.6$^{+0.8}_{-1.0}$&1.71$_{-0.04}^{+0.03}$&0.6$_{-0.4}^{+0.4}$&142/145\\ 
\tableline
\end{tabular}
\caption{Parameters best fit values of the low and high states for
comptonization model in different geometries. We also show the best fit
values obtained with {\sc pexrav} (like in Table 1, we assume in this
case $kT_e=E_{\rm c}/2$ and we compute $\tau$ as explained in section
\ref{compdiffgeom}). We assume here an inclination angle $i=30^{\circ}$.}
\label{tab2}
\end{table}

\clearpage

\pagestyle{empty}
\setcounter{figure}{0}

\begin{figure}[h]
\psfig{width=\textwidth,file=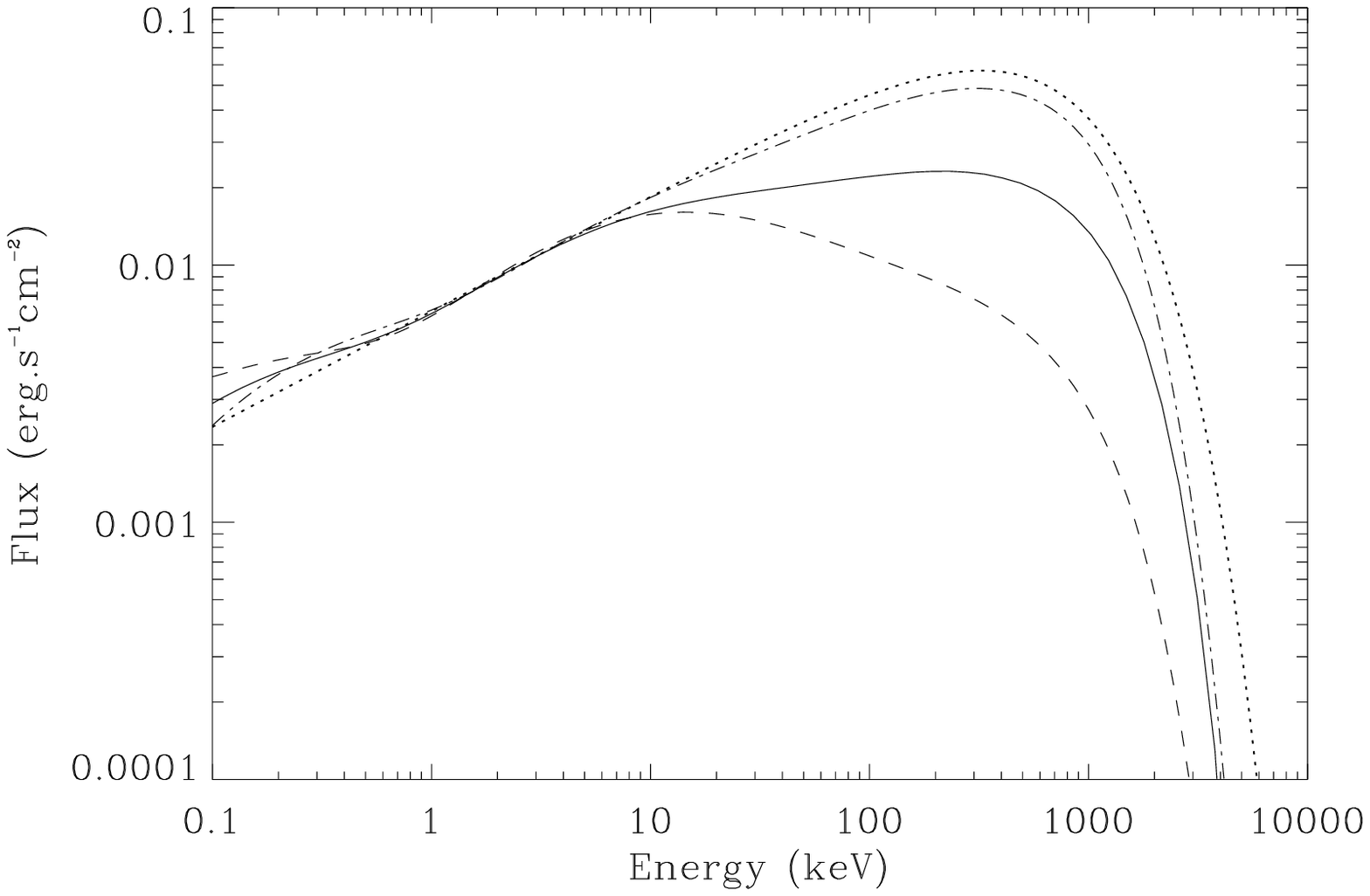,angle=0}
\caption{}
\end{figure}

\newpage

\begin{figure}[h]
\begin{tabular}{ll}
\psfig{width=\textwidth,file=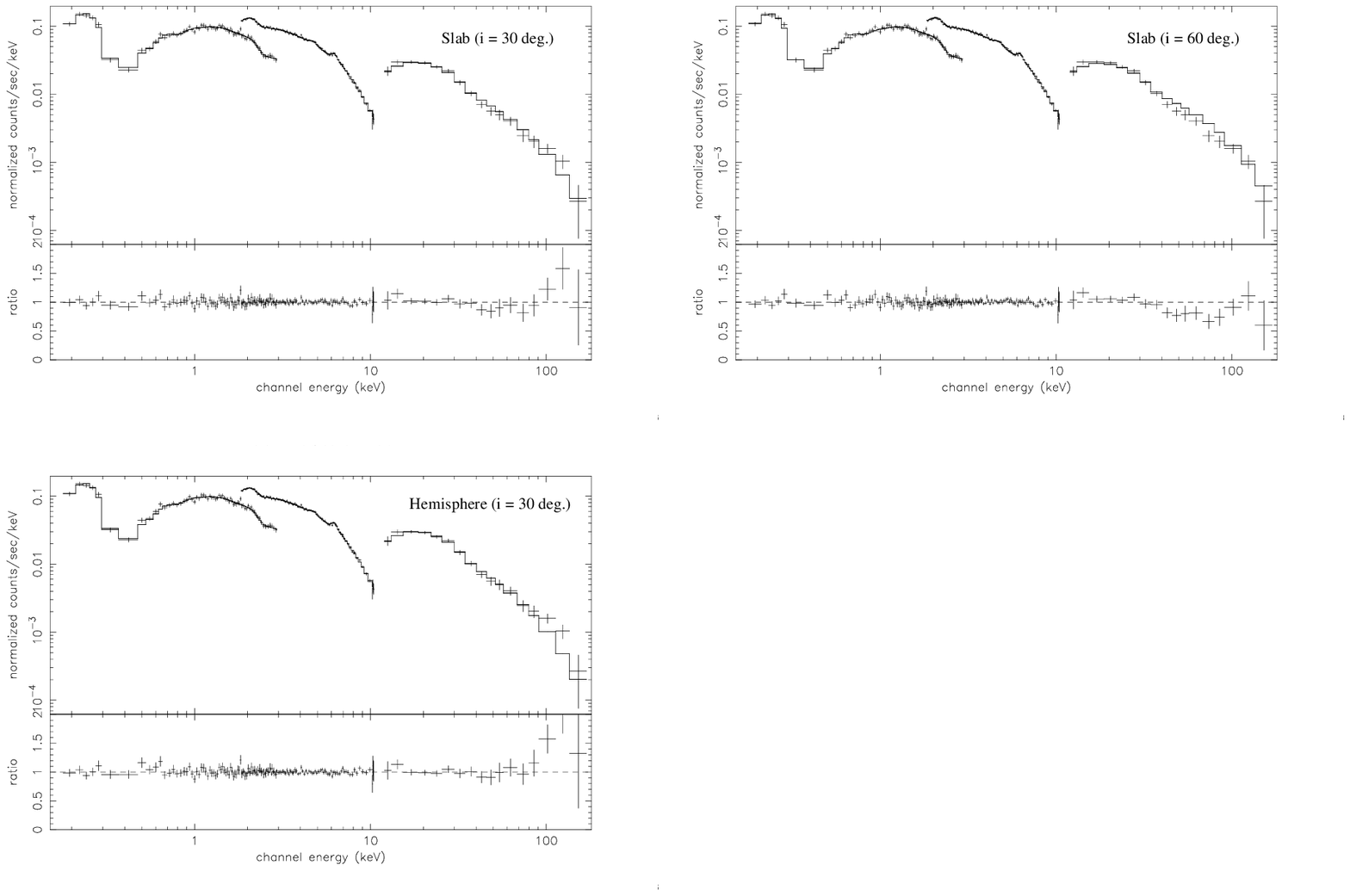,angle=0}
\end{tabular}
\caption{}
\end{figure}

\newpage

\begin{figure}
\begin{tabular}{ll}
\psfig{width=\textwidth,file=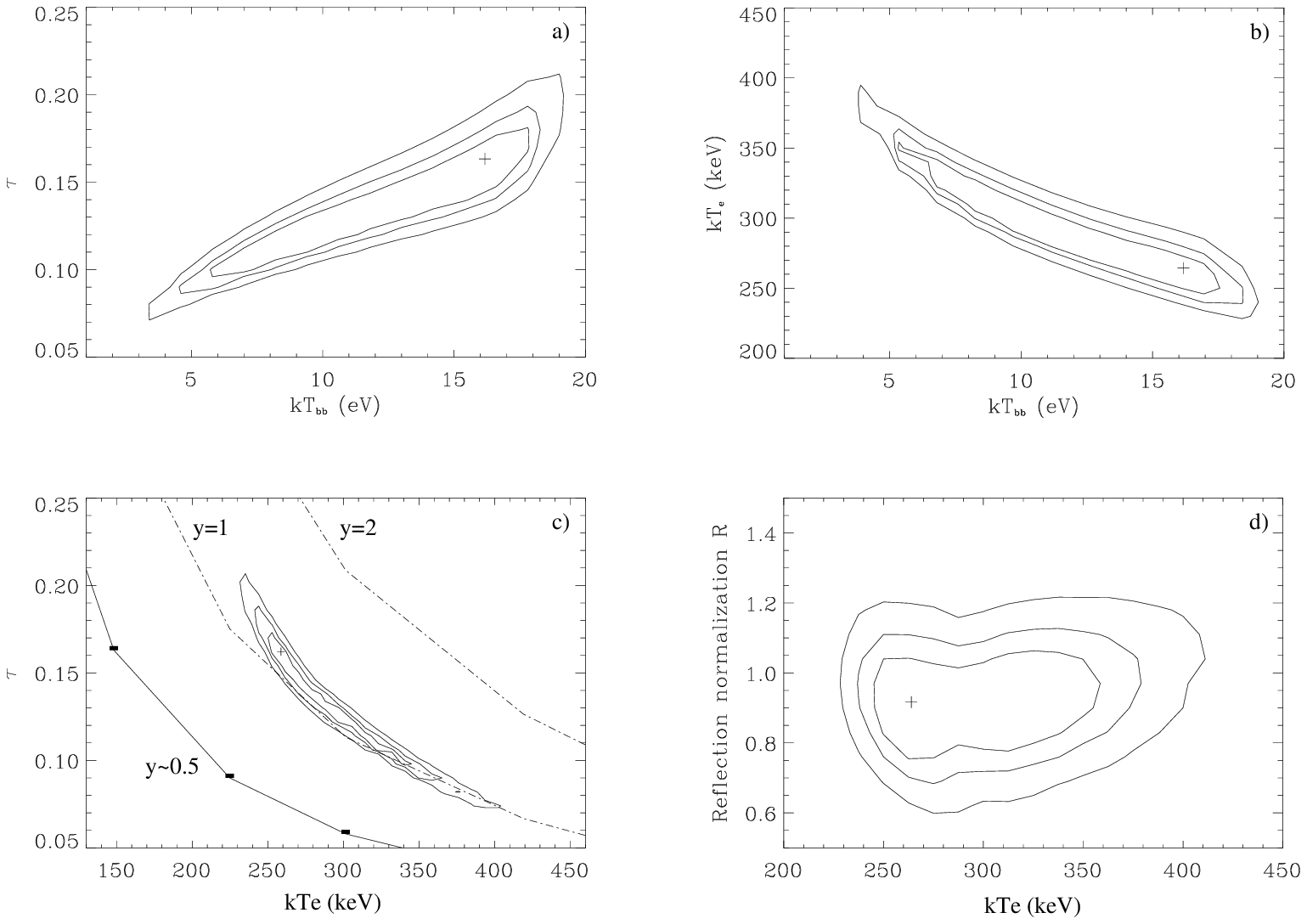,angle=0}
\end{tabular}
\caption{}
\end{figure}

\newpage
\begin{figure}
\begin{tabular}{ll}
\psfig{width=\textwidth,file=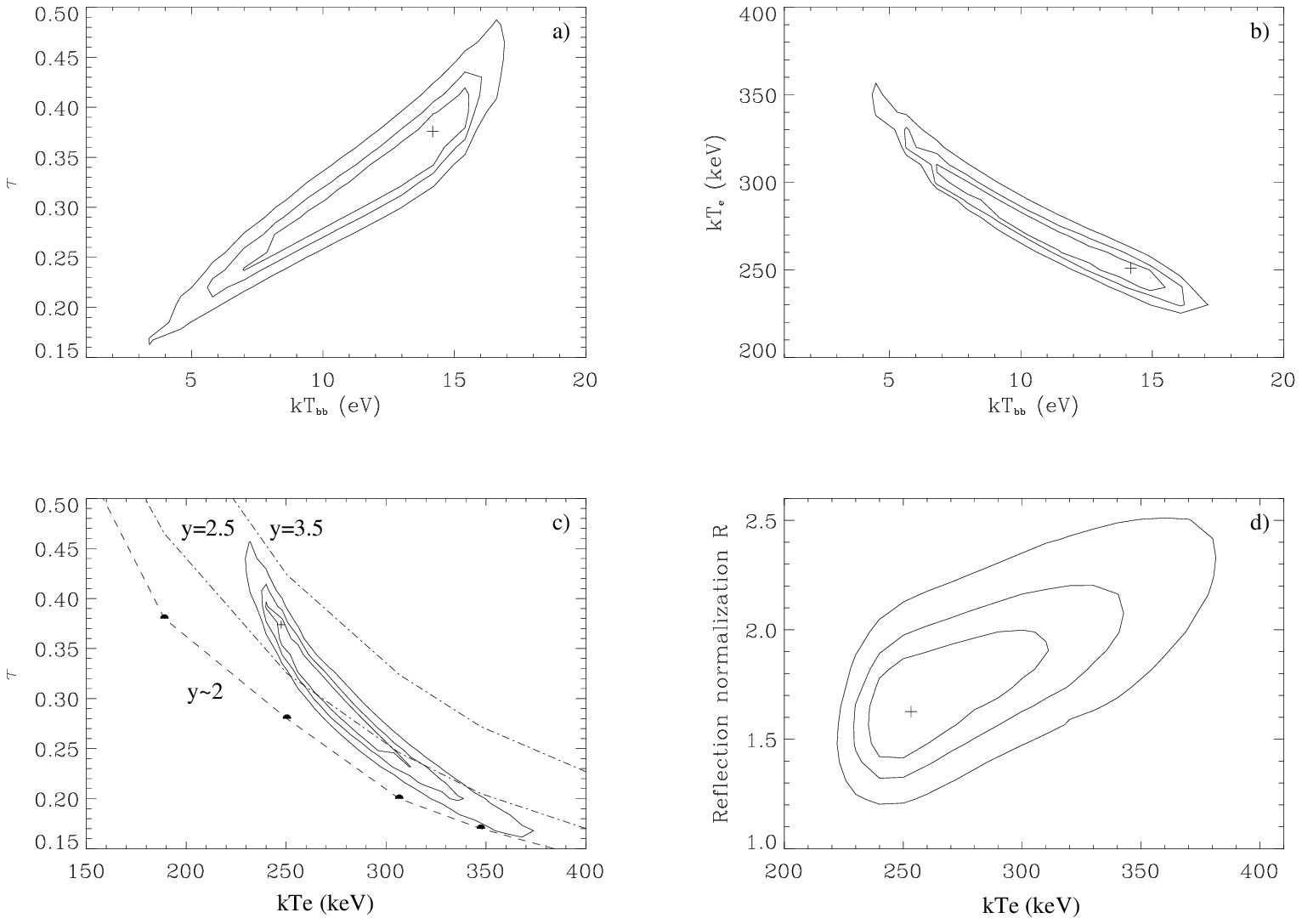,angle=0}
\end{tabular}
\caption{}
\end{figure}

\newpage

\begin{figure}[h]
\begin{tabular}{l}
\psfig{width=\textwidth,file=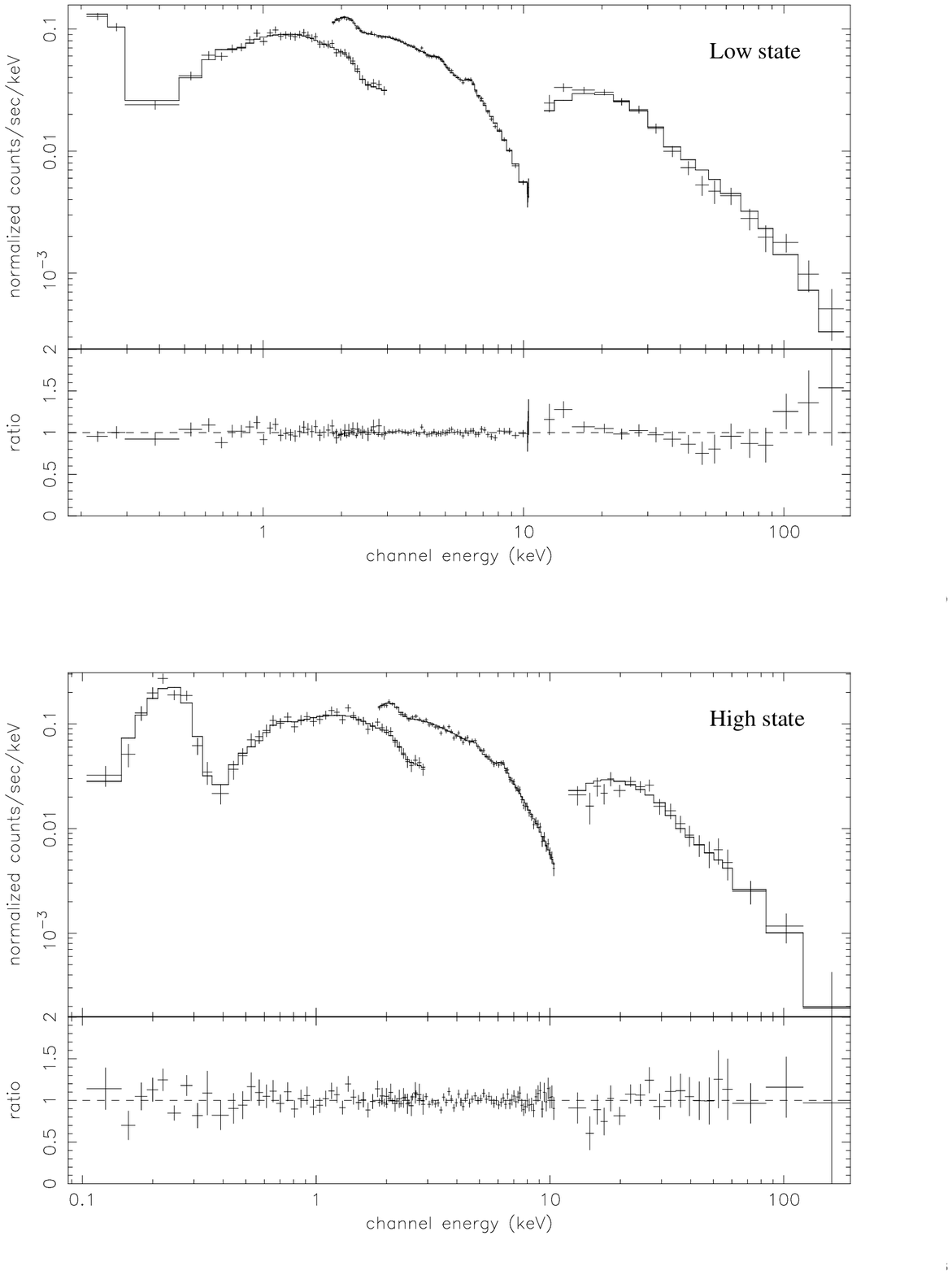,angle=0}
\end{tabular}
\caption{}
\end{figure}

\newpage

\begin{figure}[h]
\begin{tabular}{ll}
\psfig{width=\textwidth,file=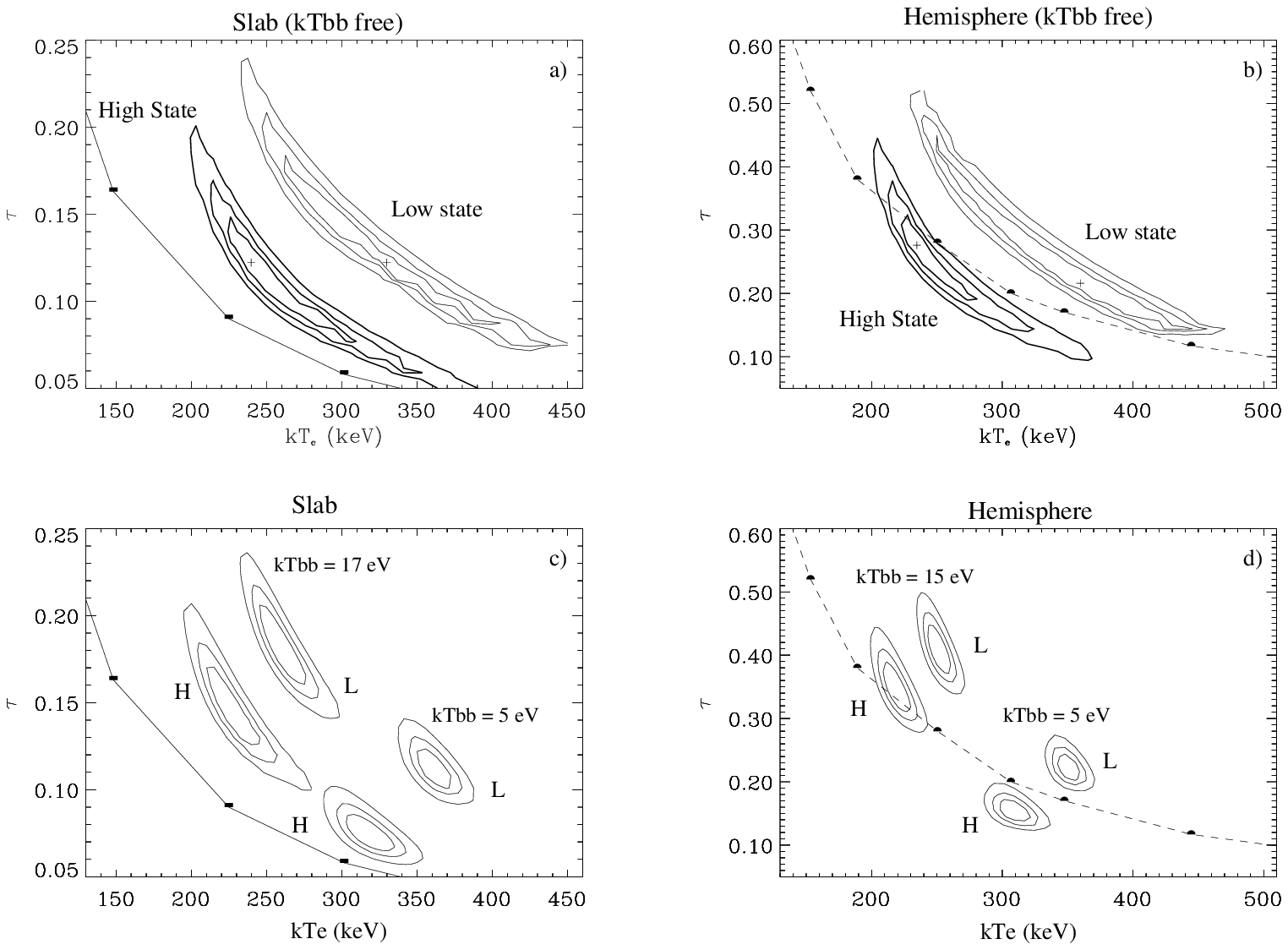,angle=0}
\end{tabular}
\caption{}
\end{figure}

\newpage

\begin{figure}
\psfig{width=\textwidth,file=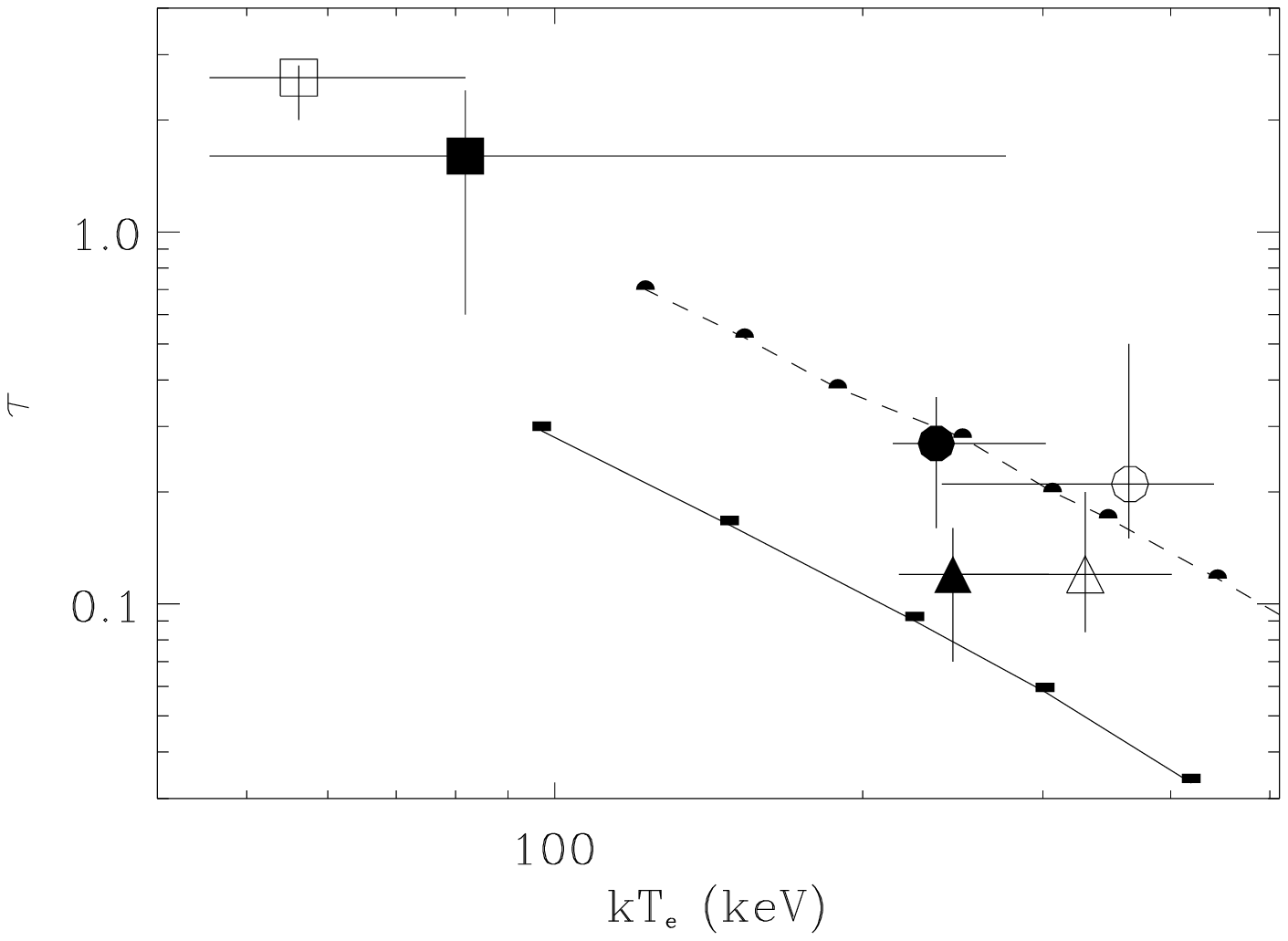,angle=0}
\caption{}
\end{figure}

\newpage

\begin{figure}[h]
\psfig{width=\textwidth,file=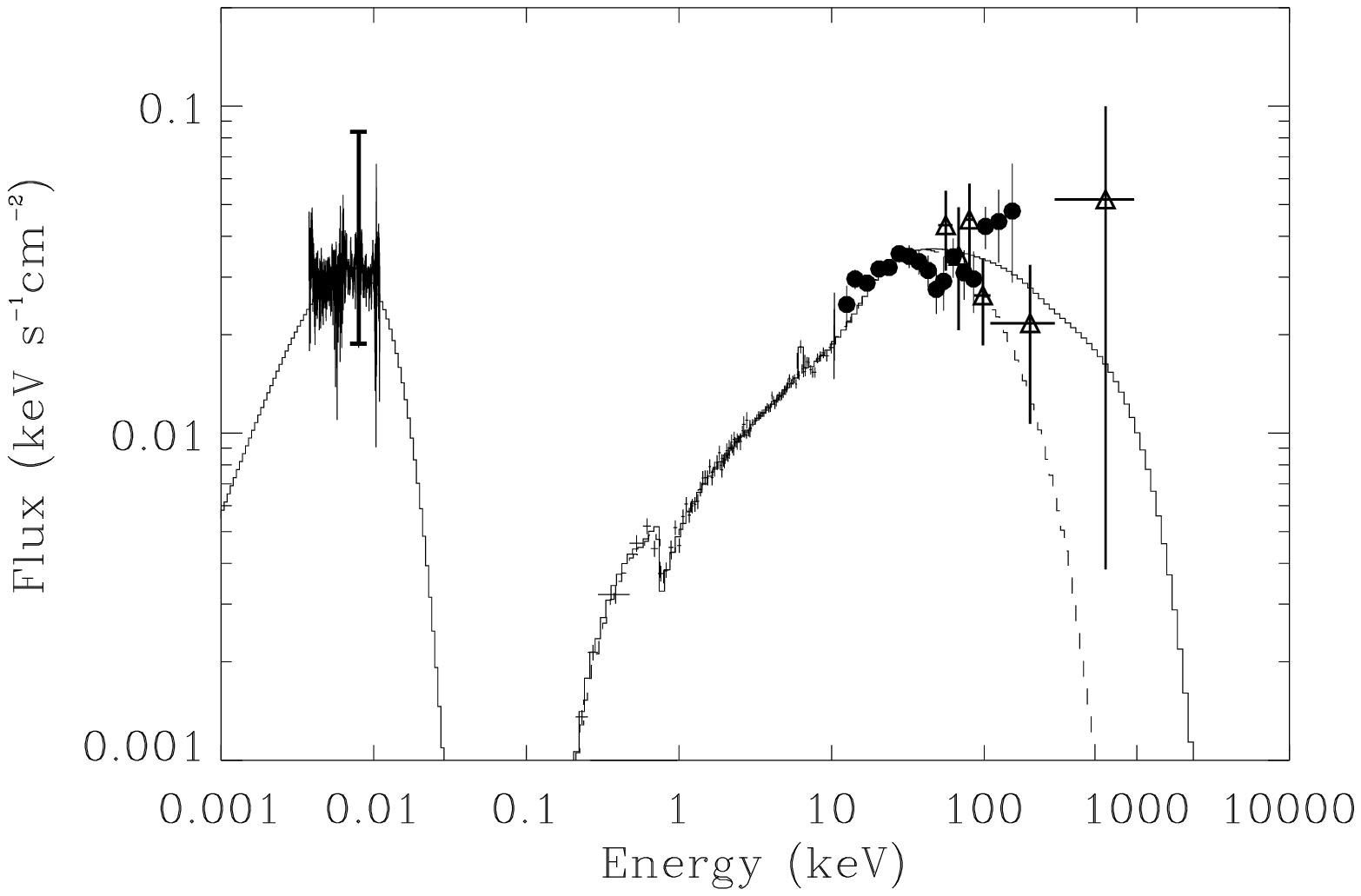,angle=0}
\caption{}
\end{figure}

\clearpage

\begin{figure}
\psfig{width=\textwidth,file=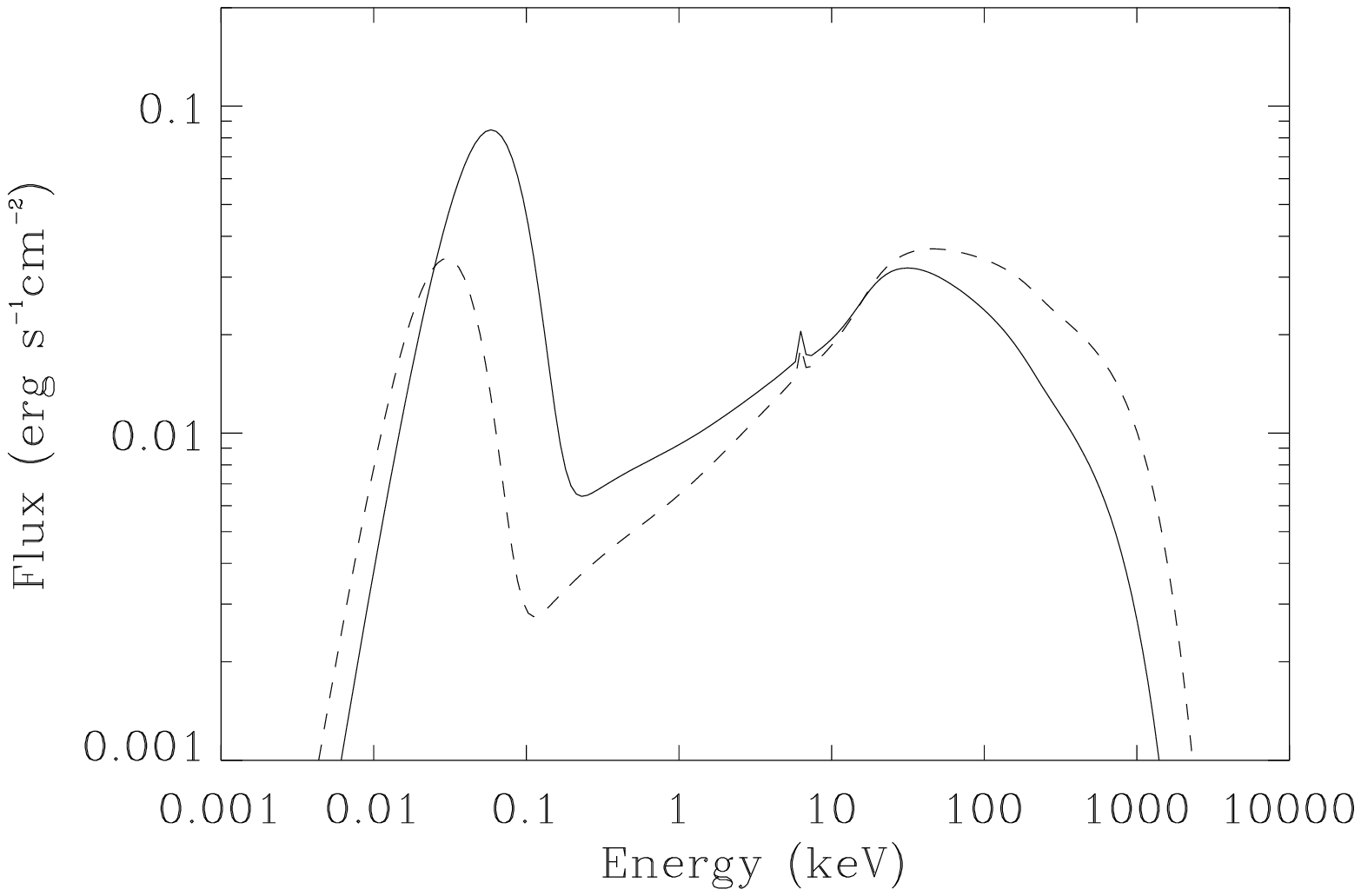,angle=0}
\caption{}
\end{figure}


\begin{thebibliography}{}
\bibitem[1996]{arn96} Arnaud, K. A. 1996, ASP Conf. 
Ser. 101: Astronomical Data Analysis Software and Systems V, 5, 17 

\bibitem[1997]{bel97} Belloni, T., Mendez, M., King, A. R., Van Der Klis,
M. \& Van Paradijs, J. 1997, \apjl, 488, L109

\bibitem[1997]{boe97} Boella, G., et al. 1997, \aaps, 122, 327

\bibitem[1992]{cla92} Clavel, J., et al. 1992, \apj, 393, 113 

\bibitem[2000]{chi00} Chiang, J., Reynolds, C. S., Blaes, O. M., Nowak,
M. A., Murray, N., Madejski, G., Marshall, H.  L. \& Magdziarz, P. 2000,
\apj, 528, 292

\bibitem[1995]{don95} Done, C., Pounds, K. A., Nandra, K. \& Fabian,
A. C. 1995, \mnras, 275, 417

\bibitem[1997]{dov97} Dove, J. B., 
Wilms, J.  \& Begelman, M. C. 1997, \apj, 487, 747 

\bibitem[1999]{fio99} Fiore, F. Guainazzi, M. and Grandi, P. 1999,
SAXabc, vs. 1.2, Cookbook for BeppoSAX NFI Spectral Analysis

\bibitem[1997]{fro97} Frontera, F., Costa, E., Dal Fiume, D., Feroci, M.,
Nicastro, L., Orlandini, M., Palazzi, E. \& Zavattini, G. 1997, \aaps,
122, 357

\bibitem[1991]{geo91} George, I. M. \& Fabian, A. C. 1991, \mnras, 249,
352

\bibitem[1994]{ghi94} Ghisellini, G.  \& Haardt, F.  1994, \apjl, 429,
L53  

\bibitem[1991]{haa91} Haardt, F. \& Maraschi, L. 1991, \apjl, 380, L51

\bibitem[1993]{haa93a} Haardt, F.  1993, \apj, 413, 680 

\bibitem[1993]{haa93b} Haardt, F.  \& 
Maraschi, L.  1993, \apj, 413, 507 

\bibitem[1994]{haa94} Haardt, F. 1994, PhD dissertation, SISSA, Trieste (H94)

\bibitem[1997]{haa97} Haardt, 
F. , Maraschi, L.  \& Ghisellini, G.  1997, \apj, 476, 620 

\bibitem[1999]{iwa99} Iwasawa K. et al., astro-ph/9904071

\bibitem[1992]{jou92} Jourdain, E., et al.  1992, \aap, 256, L38

\bibitem[1977]{lia77} Liang, E. P. T. \& Price, R. H. 1977, \apj, 218, 247

\bibitem[1987]{lig87} Lightman, A. P.  \& Zdziarski, A. A. 1987, \apj,
319, 643

\bibitem[1988]{lig88} Lightman, A. P. \& White, T. R. 1988, \apj, 335, 57

\bibitem[1995]{mag95} Magdziarz, P.  \& Zdziarski, A. A. 1995, \mnras,
273, 837

\bibitem[1998]{mag98} Magdziarz, P. , Blaes, O. M., Zdziarski, A. A.,
Johnson, W. N.  \& Smith, D. A. 1998, \mnras, 301, 179 (M98)

\bibitem[1993]{mai93} Maisack, M., et al. 
1993, \apjl, 407, L61 

\bibitem[1998]{mal98} Malzac, J., Jourdain, E., Petrucci, P. O. \& Henri,
G. 1998, \aap, 336, 807

\bibitem[1991]{mat91} Matt, G., Perola, G. C. \& Piro, L. 1991, \aap,
247, 25

\bibitem[1999]{mat99} Matt, G., Proceeding of the conference ``X-ray
Astronomy '999. Stellar Endpoints, AGN and the Diffuse Background'',
September 6-10, Bologna, Italy. To appear in Astrophysical Letters and
Communications

\bibitem[1995]{mus95} Mushotzky, R. F., Fabian, A. C., Iwasawa, K.,
Kunieda, H., Matsuoka, M., Nandra, K. \& Tanaka, Y. 1995, \mnras, 272, L9

\bibitem[1991]{nan91} Nandra, K., Pounds, K.  A., Stewart, G. C., George,
I. M., Hayashida, K., Makino, F. \& Ohashi, T.  1991, \mnras, 248, 760

\bibitem[1997]{nan97} Nandra, K., George, I.  M., Mushotzky, R. F.,
Turner, T. J. \& Yaqoob, T. 1997, \apj, 477, 602

\bibitem[1998]{nar98} Narayan, R., Mahadevan, R. \& Quataert, E. 1998,
Theory of Black Hole Accretion Disks, 148

\bibitem[1999]{nic99} Nicastro et al., 1999, ApJ, in press (N99)

\bibitem[1997]{par97} Parmar, A. N., et al.  1997, \aaps, 122, 309

\bibitem[1996]{pou96} Poutanen, J.  \& Svensson, R.  1996, \apj, 470, 249
(PS96)   

\bibitem[1976]{poz76} Pozdniakov, L. A., Sobol, I. M. \& Siuniaev,
R. A. 1976, Soviet Astronomy Letters, 2, 55

\bibitem[1995]{rey95} Reynolds, C.  S., Fabian, A. C. \& Inoue, H. 1995,
\mnras, 276, 1311

\bibitem[1997]{rey97} Reynolds, C. S. 1997, \mnras, 286, 513

\bibitem[1979]{ryb79} Rybicki, G. B. \& 
Lightman, A. P. 1979, New York, Wiley-Interscience, 1979. 393 p.,  

\bibitem[1976]{sha76} Shapiro, S.  L., Lightman, A. P. \& Eardley,
D. M. 1976, \apj, 204, 187

\bibitem[1995a]{ste95a} Stern, B. E., Poutanen, J.  , Svensson, R. ,
Sikora, M.  \& Begelman, M. C. 1995, \apjl, 449, L13

\bibitem[1995b]{ste95b} 
Stern, B. E., Begelman, M. C., Sikora, M.  \& Svensson, R.  1995, \mnras, 
272, 291 

\bibitem[1980]{sun80} Suniaev, R. A. \& 
Titarchuk, L. G. 1980, \aap, 86, 121 

\bibitem[1996]{sve96} Svensson, R. 1996, \aaps, 120, C475 

\bibitem[1994]{tit94} Titarchuk, L.  1994, \apj, 434, 570

\bibitem[1990]{wal90} Walter, R. \& Courvoisier, T. J. -L. 1990, \aap,
233, 40

\bibitem[1988]{whi88} White, T.  R., Lightman, A. P. \& Zdiziarski,
A. A. 1988, \apj, 331, 939

\bibitem[1994]{zdz94} Zdziarski, A. A. k. a. 1994, Fabian, A. C., Nandra, K., 
Celotti, A., Rees, M. J., Done, C., Coppi, P. S. \& Madejski, G. M. 1994, 
\mnras, 269, L55 

\end{thebibliography}
\end{document}